\documentclass{aa}  
\usepackage{graphicx}
\usepackage[varg]{txfonts}	
\begin{document} 



   \title{Optical--radio positional offsets for active galactic nuclei}

   \author{G. Orosz \inst{1}
         \and S. Frey \inst{2}}

   \institute{Department of Geodesy and Surveying, Budapest University of Technology and Economics, P.O. Box 91, H-1521 Budapest, Hungary\\
          \email{gabor.orosz@gmail.com}
          \and F\"OMI Satellite Geodetic Observatory, P.O. Box 585, H-1592 Budapest, Hungary\\
          \email{frey@sgo.fomi.hu}}
   \date{Received February 12, 2013; accepted Month Day, Year}

\authorrunning{}
\titlerunning{Optical--radio positional offsets for AGNs}


  \abstract
   {It will soon become possible to directly link the most accurate radio reference frame with the {\em Gaia} optical reference frame using many common extragalactic objects. It is important to know the level of coincidence between the radio and optical positions of compact active galactic nuclei (AGNs).}
   {Using the best catalogues available at present, we investigate how many AGNs with significantly large optical--radio positional offsets exist as well as the possible causes of these offsets.}
   {We performed a case study by finding optical counterparts to the International Celestial Reference Frame (ICRF2) radio sources in the Sloan Digital Sky Survey (SDSS) Data Release 9 (DR9). The ICRF2 catalogue was used as a reference because the radio positions determined by Very Long Baseline Interferometry (VLBI) observations are about two orders of magnitude more accurate than the optical positions.} 
   {We find 1297 objects in common for ICRF2 and SDSS DR9. Statistical analysis of the optical--radio differences verifies that the SDSS DR9 positions are accurate to $\sim$55 milliarcseconds (mas) in both right ascension and declination, with no systematic offset with respect to ICRF2. We find 51 sources ($\sim$4\% of the sample) for which the positional offset exceeds 170~mas ($\sim$3$\sigma$). Astrophysical explanations must exist for the majority of these outliers. There are three known strong gravitational lenses among them. Dual AGNs or recoiling supermassive black holes may also be possible.}
   {The most accurate {\em Gaia}--VLBI reference frame link will require a careful selection of a common set of objects by eliminating the outliers. On the other hand, the significant optical--radio positional non-coincidences may offer a new tool for finding e.g. gravitational lenses or dual AGN candidates. Detailed follow-up radio interferometric and optical spectroscopic observations are encouraged to investigate the outlier sources found in this study.}

   \keywords{galaxies: active --
             quasars: general --
             astrometry --
             surveys --
             catalogs --
             reference systems
             }

   \maketitle


\section{Introduction}

An ideal quasi-inertial reference system would be defined by point-like sources radiating in all wavebands of the electromagnetic spectrum, providing a simple relationship between the positions determined at different frequencies \citep{walter}. This would facilitate the cross-identification of celestial objects in the radio, infrared, optical, ultraviolet, and high-energy bands. However, in practice, reference frames at different frequencies are defined by different objects, since suitable sources usually radiate intensely only in particular wavebands (e.g. quasars can be bright in the radio but are relatively faint in the optical, while stellar objects in the Galaxy are usually bright in the optical but weak radio emitters at best). This makes it necessary to link the reference frames used in different wavebands. The link between the optical and radio domains is particularly important, since a dominant fraction of astronomical observations is made in the optical, but the most accurate reference frame is currently realized in the radio. 

The latest realization of the International Celestial Reference System (ICRS) at radio frequencies was adopted by the International Astronomical Union (IAU) in 2009. The second version of the International Celestial Reference Frame \citep[ICRF2,][]{fey2} is defined by the precise positions of selected compact extragalactic radio sources (active galactic nuclei, AGNs) regularly observed with Very Long Baseline Interferometry (VLBI) over a long period of time.

Links between the ICRF2 and optical reference frames can be either direct or indirect. A direct connection would mean that the positions of the radio AGNs are directly measured in the optical band. At present, however, the most accurate optical astrometric catalogues cannot be linked directly to the ICRF2, because their limiting magnitude significantly exceeds the detection limit of AGNs in the optical. For example, in the Sloan Digital Sky Survey (SDSS) Quasar Catalogue \citep{paris}, $m_{\rm q}$$\gtrsim$$17\,{\rm mag}$. On the other hand, the primary astrometric reference used in the optical domain, the Hipparcos catalogue \citep{perryman} has a limiting magnitude of only $V$$\approx$12.4$\,{\rm mag}$. It is defined by bright stellar sources and could only be connected to the ICRF indirectly \citep{kovalevsky1,stone}. Since no extragalactic radio source could be reliably detected with Hipparcos, a variety of secondary methods had to be used, e.g. relative astrometric measurements of radio stars with respect to nearby reference quasars with VLBI and connected-element radio interferometers, sensitive optical observations of quasars relative to Hipparcos stars, and a comparison of VLBI-determined and optically determined Earth orientation parameters. The problem with these solutions is that since the indirect links use stellar sources for the connection between the radio and the optical reference frames, the quality of the link degrades with time due to the uncertainties in the measured stellar proper motions.

However, with the sensitive next-generation space astrometry mission, the European Space Agency's {\em Gaia} spacecraft \citep[e.g.][]{mignard,mccaughrean} to be launched in the second half of 2013, a quasi-inertial reference frame can directly be established by around 2020, based on measurements of a large number of extragalactic sources (probably tens of thousands of primary objects) in the optical as well. 
{\em Gaia} will detect a total of $\sim$500\,000 quasars brighter than the limiting magnitude of 20, with a precision similar to ICRF2 \citep[$\leq$0.1 milliarcseconds, mas,][]{charlot}. It will become possible to directly link the radio and optical reference frames using a large number of common objects for the first time. For consistency, it is important to make the alignment with the highest possible accuracy, which requires common objects with excellent optical and radio astrometric properties \citep[e.g.][]{fey1}. As detailed by \citet{bourda1, bourda2, bourda3}, additional observations of new radio sources and the construction of the ICRF3 are needed for the best possible link, since the potentially most suitable {\em Gaia} sources for the alignment will not necessarily be the best ICRF2 sources. The effect of secular aberration drift, i.e. the apparent proper motion of extragalactic sources caused by the rotation of the solar system around the Galactic centre \citep{kovalevsky2, titov}, also has to be taken into account when constructing the next ICRF \citep{liu}.

The direct connection between the {\em Gaia} celestial reference frame and the ICRF is essential not only for astrometry, but for astrophysical reasons as well. It will become possible to accurately study the coincidence between the radio and optical emission peaks of AGNs on a sub-mas scale \citep[e.g. core shift,][]{lobanov,kovalev}, and the observations of the AGNs at different wavelengths will only be interpreted correctly if the measurements are expressed in a consistent system \citep[e.g.][]{ivezic, kimball}.

While waiting for the {\em Gaia} extragalactic reference frame to be constructed, we can perform case studies for the direct reference frame link using currently available large optical sky surveys. The SDSS \citep{york} is currently the largest sky survey available, and although not an astrometric catalogue, the faint limiting magnitude ($V$$\approx$22$\,{\rm mag}$ for 95\% completion) makes it possible to identify the counterparts of many radio-loud AGNs that have accurate radio positions. There have been several studies cross-referencing the SDSS optical positions directly with the radio positions in the ICRF \citep{frey1, lambert, souchay1, damljanovic}, or comparing other radio and optical properties based on a common optical--radio sample \citep{ivezic, kimball} using e.g. the Very Large Array (VLA) Faint Images of the Radio Sky at Twenty-centimeters (FIRST) survey catalogue \citep{becker}. Since the astrometric reductions of SDSS can be traced back to the Hipparcos catalogue, it can also be used to locally connect the radio and optical reference frames directly by finding common sources in the SDSS and radio catalogues. The ICRF positions of these optical--radio pairs are then determined precisely through relative VLBI astrometry using nearby ICRF sources as phase-reference calibrators \citep[e.g.][]{frey3}. In addition, there are other optical catalogues that are sensitive enough to make direct connection of quasar positions possible between the radio and the optical \citep[e.g.][]{assafin, souchay1}, such as the US Naval Observatory (USNO) CCD Astrograph Catalogue \citep[UCAC3,][]{zacharias2}, the 2-degree Field Quasar Redshift Survey \citep[2QZ,][]{croom}, or other quasar catalogues compiled using the above-mentioned radio and optical surveys, e.g. the Large Quasar Astrometric Catalogue \citep[LQAC-2,][]{souchay2} or the \citet{veroncetty} catalogue. Finally, these studies not only provide an independent assessment of the astrometric accuracy of the optical surveys \citep[e.g.][]{souchay1}, but the comparison of the optical and radio positions could reveal some new information about the physical properties of these AGNs.

In this paper, we present a case study of a direct astrometic link between the radio and the optical bands by comparing the VLBI positions of AGNs in the ICRF2 with those in the 9$^{\rm th}$ Data Release (DR9) of the SDSS \citep{ahn}. The astrometric properties of the two catalogues are reviewed and our sample selection process is explained in Sect.~\ref{sec:crossref}. Section~\ref{sec:statistics} gives the statistical characterization of the optical--radio coordinate differences. We find a large number of sources that are significantly (>3$\sigma$) offset between the radio and the optical. These sources are presented in Sect.~\ref{sec:outliers}. Possible reasons behind the ``outliers'' are discussed in Sect.~\ref{sec:discussion}, including astrometric errors and various potential astrophysical explanations. Conclusions are drawn in Sect.~\ref{sec:conclusion}.


\section{Selecting common sources in ICRF2 and SDSS DR9}
\label{sec:crossref}

\subsection{Astrometric properties of the SDSS Data Releases}
\label{subsec:sdss}

The SDSS DR9 catalogue{\footnote {\url{http://www.sdss3.org/dr9/}} \citep{ahn} is the newest release of the SDSS-III campaign \citep{eisenstein}, which is an extension of the previous SDSS-I and SDSS-II projects \citep{york}. It covers $\sim$14\,500 square degrees of the sky in the optical, mainly in the northern galactic hemisphere between right ascension 7$^{\rm h}$$<$$\,\alpha\,$$<$18$^{\rm h}$ and declination $-5\degr$$<$$\,\delta\,$$<$$+70\degr$. It also covers a smaller region in the southern galactic hemisphere between 21$^{\rm h}$$<$$\,\alpha\,$$<$4$^{\rm h}$ and $-15\degr$$<$$\,\delta\,$$<$$+35\degr$. The large sky coverage and the faint ($V$$\approx$$22\,{\rm mag}$) limiting magnitude make it possible to identify the counterparts of many radio-loud AGNs that have accurate radio positions available in the ICRF2. The astrometric calibration of SDSS is described in detail by \citet{pier}. The source positions derived from the $r$ photometric CCDs ($\lambda_{\rm r}$=6165\AA) are calibrated using the 2$^{\rm nd}$ release of the USNO CCD Astrograph Catalogue \citep[UCAC2,][]{zacharias1} and the UCAC r14 catalogue (a supplemental set of UCAC at declinations above $41\degr$). Proper motions are derived from the SDSS+USNO-B catalogue \citep{munn}. There are $\sim$2$-$3 magnitudes of overlap between UCAC and unsaturated stars on the $r$ photometric CCDs. The UCAC observations are based on the Tycho-2 reference stars \citep{hog}, which are directly connected to the Hipparcos Reference Frame (HRF). \citet{pier} declares that the global astrometric precision of SDSS relative to HRF is 45~mas rms per coordinate at $r$$\approx$$20\,{\rm mag}$ and approximately 100~mas rms at $r$$\approx$22$\,{\rm mag}$, with additional systematic errors of less than 20~mas resulting from the reference catalogues.

Since DR9 is currently the newest, the most accurate, and has the largest coverage among the SDSS data releases, we used it for cross-referencing with the VLBI-measured reference positions in ICRF2. It also includes the first-year data of the new Baryon Oscillation Spectroscopic Survey (BOSS) spectographs \citep{dawson}, which focus on obtaining new spectra of galaxies in the redshift range 0.15$<$$z$$<$0.8 and quasars with 2.15$<$$z$$<$3.5, and have a much denser and larger coverage than the previous spectroscopic surveys of SDSS-I and SDSS-II described by \citet{abazajian}.

\subsection{Astrometry of ICRF2}
\label{subsec:icrf2}

The fundamental celestial reference system adopted by the IAU has been based on the radio coordinates of AGNs since 1998 \citep{feissel}. It was first realized by ICRF1 \citep{ma} based on the radio positions of 212 extragalactic sources distributed over the entire sky. Due to the accumulating observational data and significant developments and improvements in astrometric VLBI sensitivity and quality, ICRF was redefined by the International Earth Rotation and Reference Systems Service (IERS) and the International VLBI Service for Geodesy and Astrometry (IVS) in 2009. The resulting ICRF2{\footnote {\url{http://hpiers.obspm.fr/icrs-pc/}} is currently the realization of the celestial reference system at the radio frequencies and is described in detail by \citet{fey2}. It contains the precise positions of 3414 compact radio sources, measured with the VLBI technique, with a noise floor of only 40 microarcseconds ($\mu$as) and an axis stability of 10~$\mu$as. The coordinate system is maintained using a set of 295 defining sources selected on the basis of positional stability and the lack of extensive intrinsic source structure. Of the total number of sources, 2197 were observed only in the Very Long Baseline Array (VLBA) Calibrator Survey \citep[VCS,][and references therein]{petrov}, most of them in only one VCS session. These sources are located at $\delta$$>$$-30\degr$ and their coordinates in most cases are as accurate as those of the non-VCS sources, which cover the whole sky uniformly. The precision of the source coordinates in ICRF2 is better than 1~mas in both right ascension and declination, ranging between $\sim$0.1$-$0.5~mas depending on the different surveys and the number of observations \citep{fey2}. 

\begin{figure}[]
\includegraphics[width=9.0cm, clip=]{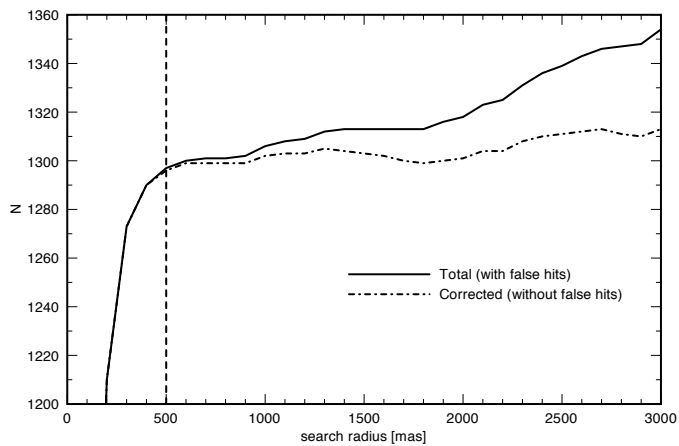}
\caption{
Graph of the number of counterparts to ICRF2 sources found in SDSS DR9 as a function of the search radius. The continuous curve represents the total number of hits ($N$), whereas the dot-dashed curve shows the results corrected with the false match ratio derived from Monte Carlo simulation ($N$$-$$p_{\rm f}N$) detailed in Sect.~\ref{subsec:sourceselection}. The growing deviation of the two curves is caused by the increase of false identifications at larger search radii. The curves nearly coincide up to 500~mas search radius, the value used in this case study, therefore the number of false optical--radio identifications is negligible.
}
\label{img:radius}
\end{figure}

\begin{figure}[]
\includegraphics[width=9.0cm, clip=]{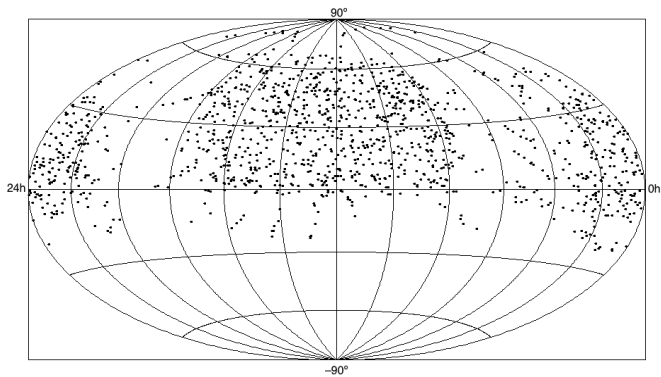}
\caption{
Sky plot (equatorial coordinates in Aitoff projection) of the 1297 AGNs found in the SDSS DR9 based on a 500-mas radius search using the ICRF2 radio source catalogue as reference. This is the sample used in our case study of the direct link between the radio and optical reference frames. The objects are distributed uniformly in the area covered by SDSS DR9.
}
\label{img:skyplot}
\end{figure}

\begin{figure*}[]
\centering
\includegraphics[bb = 0 11 478 450, width=8.5cm, clip=]{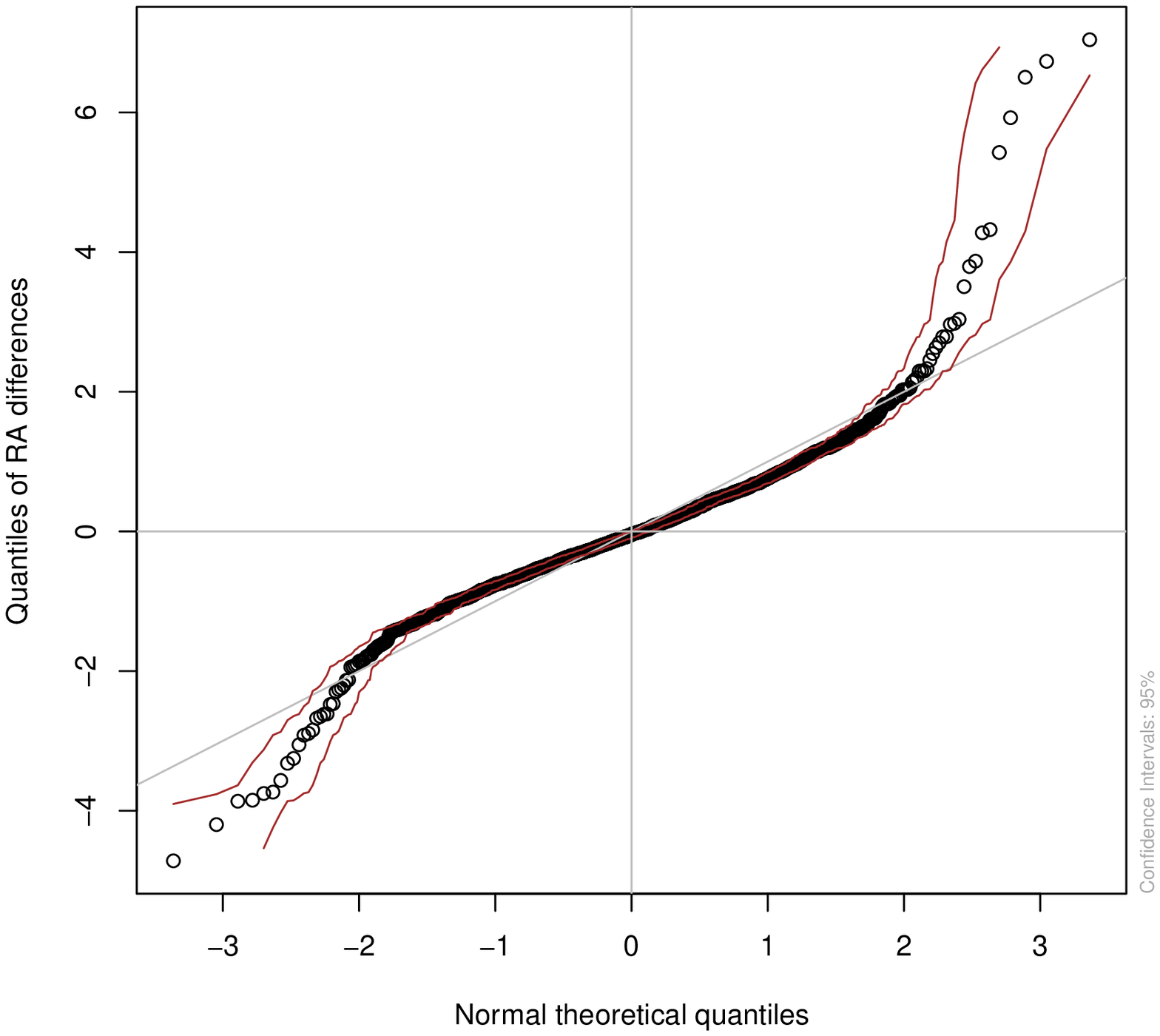}
\includegraphics[bb = 0 11 478 450, width=8.5cm, clip=]{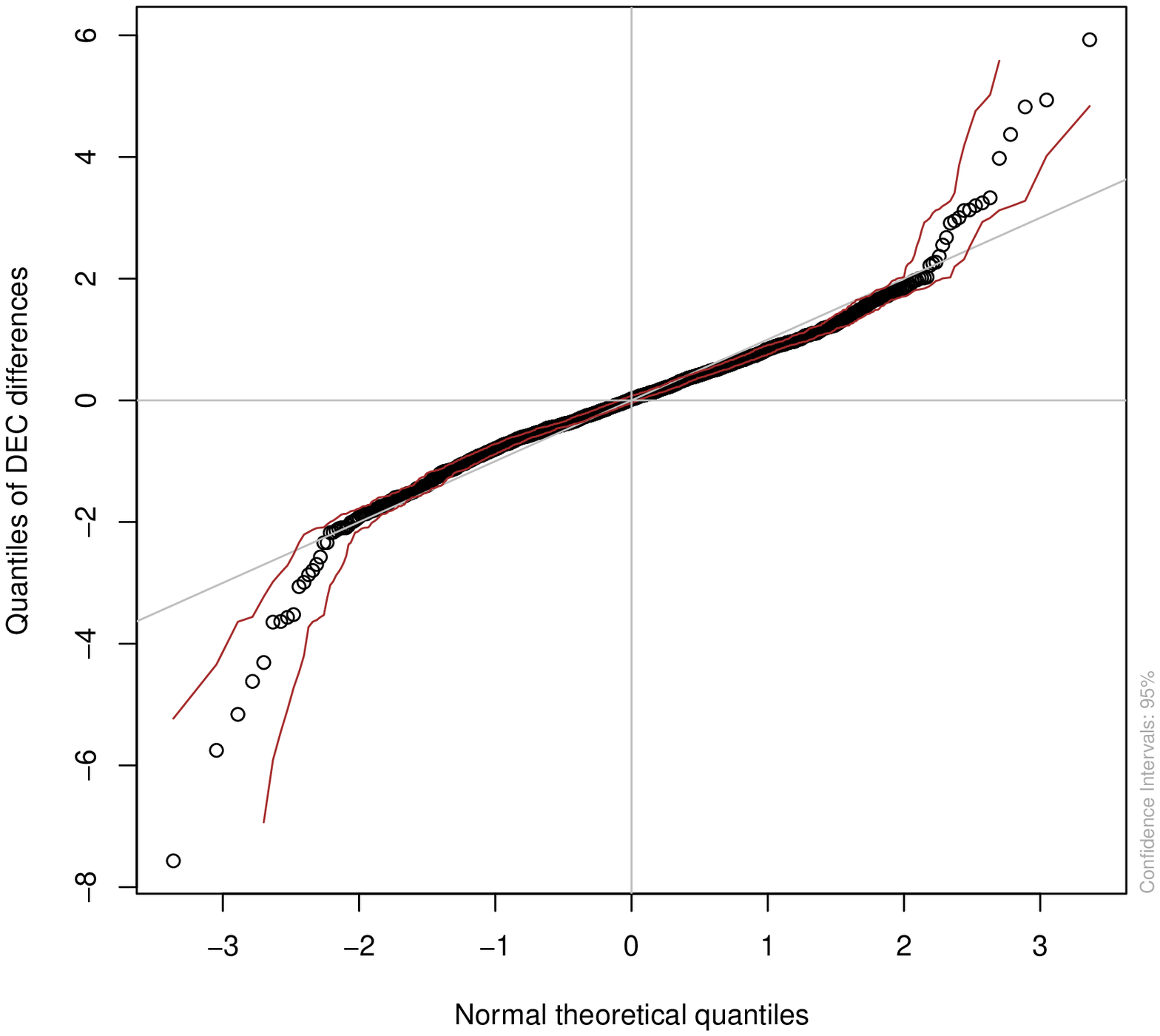}
\caption{
Normal probability plots of the right ascension (left) and declination (right) optical--radio coordinate differences. The data are normally distributed because the graphs approximately coincide with the $45\degr$ reference line. There are considerably more objects at both ends of the ordered sample data, however, which indicates a heavy tail in the distribution. The curves surrounding the quantile points indicate the 95\% confidence interval. The quantiles of the sample data are normalized to the same length reference distribution of $N(0,1)$.
}
\label{img:qqplot}
\end{figure*}

\subsection{Source selection}
\label{subsec:sourceselection}

Since the astrometric precision of ICRF2 is more than two orders of magnitude better than that of the SDSS, we considered the radio coordinates as the accurate positions of the AGNs when selecting their optical counterparts from the SDSS catalogues. This selection was accomplished by defining a search radius around the supposedly ``error-free'' radio positions and identifying the optical sources inside these circles. This selection method has been used several times in the past with various cut-off radii, between 300$-$1000~mas, when comparing ICRF--SDSS positions \citep[e.g.][]{frey1, frey2, lambert, souchay1} and $1\farcs5$$-$$3\arcsec$ when using the FIRST radio catalogue \citep[e.g.][]{ivezic}. The value chosen for the search radius matters because a smaller-than-ideal radius excludes sources that would otherwise have optical counterparts. A larger-than-ideal radius would contaminate the sample with chance identifications, i.e. sources that are in close proximity to the AGNs on the celestial sphere, but have no physical relation to them.

As a starting point, the size of the cut-off should be at least as big as the $\sim$2--3$\sigma$ astrometric precision of the less accurate catalogue, in our case $\sim$200~mas. To determine the ideal search radius, we performed a Monte Carlo simulation to calculate the probability of chance coincidences (i.e. false identifications) as a function of the radius used. We constructed eight false radio source lists by simply shifting $\alpha$ and $\delta$ for all ICRF2 sources by large arbitrary amounts, $+(1, 2, 3, 4)\degr$. We then tried to find SDSS optical counterparts for these fake ``objects'' (over 27\,000 in total) using various search radii ranging from 200$-$3000~mas, with 100~mas increments. The simulation showed that the probability of chance coincidences ($p_{\rm f}$) in DR9 is $\sim$0.01\% for 200~mas, remains $<$0.1\% up to 600~mas, and reaches $\sim$3\% at 3000~mas. The simulations for SDSS DR7 and DR8 produced similar results and generally agree well with \citet{ivezic} and \citet{souchay1}, who compared other catalogues. To find the cut-off radius to use for this case study, we also searched for optical counterparts around the real ICRF2 positions, and determined the $N$ number of matches as a function of the radius. Using the previously obtained $p_{\rm f}$ probabilities, we subtracted from $N$ the number of false matches, $p_{\rm f}N$. This provided a count corrected for the contaminating false identifications. Figure~\ref{img:radius} shows the number of total hits ($N$) and the corrected hit count ($N$$-$$p_{\rm f}N$) as a function of the search radius. This indicates that the ideal cut-off radius is around 500~mas, since the sample of the optical--radio matches is the largest here without many chance coincidences.

Among the all-sky set of 3414 ICRF2 sources, optical counterparts of 1297 ($\sim$38\%) were found in the SDSS DR9 within the search radius of 500~mas (Fig.~\ref{img:skyplot}). This is consistent with the $\sim$35\% sky coverage of the optical catalogue and indicates that practically all of the ICRF2 radio AGNs do have an optical counterpart with the SDSS limiting magnitude. The identified objects are evenly dispersed within the region covered by DR9 (see Sect.~\ref{subsec:sdss}). In the sample data set, the probability of chance coincidences is only $\sim$0.06\%, i.e. less than 1 false optical identification is expected. This provides us with a clean optical--radio AGN sample since these optical counterparts are real identifications. Among the 1297 sources, 233 are classified as extended (i.e. galaxies) and 1064 as point-like (i.e. quasars) in the SDSS DR9. All are primary objects, i.e. their position is from the best run in case of multiple observations. The optical coordinates are derived from the $r$ photometric CCDs, their average apparent magnitude is $r$$\approx$18.9$\,{\rm mag}$. Using this sample of common sources in the ICRF2 and SDSS DR9 catalogues, we determine and analyse the offsets between the optical and radio positions in Sect.~\ref{sec:statistics}.


\section{Statistical characterization of optical--radio positional differences}
\label{sec:statistics}

Using the sample data set of the 1297 optical--radio $(\alpha,\delta)$ coordinate pairs, we calculated the angular differences between the SDSS DR9 and ICRF2 positions. As mentioned in Sect.~\ref{subsec:icrf2}, the radio positions can be taken as the reference for the AGNs, thus the optical [1] minus radio [2] coordinate differences can be calculated using spherical trigonometry formulae as

\begin{equation}
  \Delta\alpha = (\alpha_{1} - \alpha_{2}) \cos\delta_{2} \textrm{ ,}
\end{equation}
\begin{equation}
  \Delta\delta = \delta_{1} - \delta_{2} \textrm{ ,}
\end{equation}
\begin{equation}
  \cos\Delta=\cos(\alpha_{1}-\alpha_{2})\cos\delta_{1}\cos\delta_{2}+\sin\delta_{1}\sin\delta_{2} \textrm{ ,}
\end{equation}
\noindent
where $\Delta\alpha$ and $\Delta\delta$ are the differences in right ascension and declination, respectively, and $\Delta$ is the total angular difference between the optical and radio positions.

\subsection{Testing of normality}

To characterize statistically the optical--radio positional differences, we tested the normality of the calculated random variables. This was accomplished by constructing a normal quantile--quantile (Q-Q) plot for both $\Delta\alpha$ and $\Delta\delta$ (Fig.~\ref{img:qqplot}). Normal Q--Q plots are probability plots that compare a sample data of unknown distribution with the standard normal distribution $N(0,1)$. A Q--Q plot is commonly used for comparing a data set to a theoretical model and is a robust graphical method of distribution analysis \citep[e.g.][]{fisher, evans, rosenkrantz, das}. It is also used in astronomical data interpretation \citep[e.g.][]{konig, pestana, huff}. In our study, the general trend of the points is somewhat flatter in both cases than the 45$\degr$ line, which indicates that the middle part of the data is slightly less dispersed than an $N(0,1)$ distribution. Moreover, as seen from the arcs at both ends on the probability plots (Fig.~\ref{img:qqplot}), the distributions of our data have heavier tails than a normal distribution. However, since the quantile pairs on both normal Q--Q plots approximately lie along the line $y=x$, both the right ascension and declination differences can be considered to have a normal distribution with a fat tail. The correlation coefficient between the two quantities is very low, 0.017. This means that $\Delta\alpha$ and $\Delta\delta$ can, for all things considered, be treated as independent normal random variables. Because the involved angular distances are very small, the total positional offset can be approximated with planar trigonometry as $\Delta$$\approx$$\sqrt{\Delta\alpha^2+\Delta\delta^2}$ , so we can assume a Rayleigh distribution for its statistical characterization.

\subsection{Data weighting and statistical parameters}

\begin{figure}[]
\includegraphics[width=9.0cm, clip=]{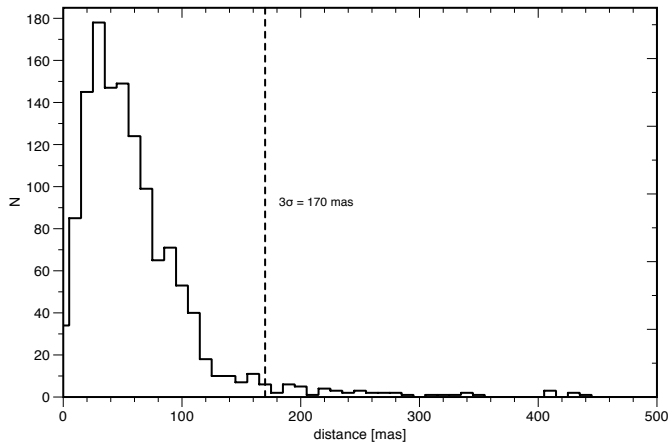}
\caption{
Histogram of optical--VLBI total positional differences for 1297 ICRF2 sources identified in the SDSS DR9 within a search radius of 500~mas. The distribution is consistent with $\sim$57~mas (1$\sigma$), assuming Rayleigh distribution. The vertical dashed line indicates the 3$\sigma$ value of 170~mas. A total of 51 sources have optical--radio offsets larger than 3$\sigma$.  
}
\label{img:histogram}
\end{figure}

When determining the Gaussian parameters of $\Delta\alpha$ and $\Delta\delta$, it is possible to refine the calculations by introducing weights characterizing the reliability of the optical positions. SDSS uses algorithms \citep[described in][]{petrosian,blanton,yasuda} to automatically distinguish extended sources (i.e. galaxies) from point-like objects (i.e. stars or quasars). However, as detailed in \citet{ahn}, some inconsistencies in classification have been found in earlier SDSS Data Releases. To avoid any possible problem, we did not rely on the galaxy/quasar classification of SDSS  when calculating the statistical characteristics of the offsets between the radio and optical positions. Instead, we took the Petrosian radius \citep{blanton,yasuda}, a measure of the optical size of a given source, and used it as a reciprocal weight for the individual source coordinates. This way we could take into account that point-like sources have more reliable position measurements. At the same time, we did not exclude extended optical sources as in an earlier study with ICRF2 and SDSS DR7 \citep{orosz}.

The distribution of the $\Delta$ positional differences of SDSS--ICRF2 AGNs is shown in the histogram of Fig.~\ref{img:histogram}. It is generally consistent with a 57-mas positional uncertainty, very close to the declared global SDSS astrometric precision \citep{pier}. The weighted optical minus radio right ascension differences ($\Delta\alpha$) have $\sigma_{\Delta\alpha}$$=$$55$~mas standard deviation and a negligible $\mu_{\Delta\alpha}$$=$$-3$~mas mean value. The weighted declination differences ($\Delta\delta$) have similar standard deviation of $\sigma_{\Delta\delta}$$=$$54$~mas and also a negligible $\mu_{\Delta\delta}=4$~mas mean value. As found in the normality test, the distribution has a fat tail, which means that there are significantly more sources with $>$3$\sigma$ positional offset than expected statistically. The theoretical distribution would only give $\sim$four sources (0.27\%) above the 3$\sigma$ level, as opposed to the detected 51 outliers, i.e. AGNs with optical--radio separations exceeding 170~mas. They represent $\sim$4\% of the sample. Furthermore, $\sim$1\% of the sources are even beyond the 5$\sigma$ mark, which should be practically impossible in a population with standard normal distribution. Earlier studies using SDSS DR4 data with 524 matching quasars and galaxies \citep{frey1}, SDSS DR5 data with 735 matching objects \citep{frey2}, and SDSS DR7 data with 806 matching quasars \citep{orosz} led to similar results, with outlier ratios also around 4\%.


\section{Sample of positional outlier objects}
\label{sec:outliers}

Comparing the radio positions of ICRF2 sources to their optical counterparts in SDSS DR9, we found a total of 51 AGNs where the optical--radio offset exceeds the 3$\sigma$ level determined in Sect.~\ref{sec:statistics}. These sources apparently show an even distribution in the sky within the coverage of SDSS DR9. Except for J1506+3730, none of them is a defining object of ICRF2. The complete list of positional outliers is presented in Table~\ref{table:outliers}. For each AGN, Col. 1 gives a short IAU designation constructed from the J2000.0 equatorial coordinates as JHHMM$\pm$DDMM. Note that the format of the complete ICRF2 designations that can be derived from the right ascension and declination is ICRF JHHMMSS.s$\pm$DDMMSS \citep{fey2}.
Based on information from the literature, in Sect.~\ref{subsec:individualsources} we comment on the properties of seven objects marked here with asterisks.
The J2000.0 equatorial coordinates are taken from the ICRF2 catalogue \citep{fey2} and listed in Cols.~2 and 3. The total optical--radio angular offsets ($\Delta$), and the offsets broken down to coordinate components ($\Delta\alpha$ and $\Delta\delta$) are given in Cols.~4, 5, and 6, respectively. The apparent $r$ magnitudes ($m_{\rm r}$) taken from SDSS DR9 can be found in Col.~7. Redshifts (where available) and a simple optical classification from SDSS DR9 (Q: quasar, G: galaxy) are given in Cols.~8 and 9. Column~10 provides the references for redshifts.

We found spectroscopic redshift measurements in the literature for 27 out of 51 sources. These redshifts range between 0.04$\la$$z$$\la$3; half of them are below 0.6, and about a quarter of the sources have $z$$\ga$2. We examined whether there is a connection between the angular separation of the optical--radio positions and the redshifts, and found no evidence for it. About one third of the outliers are classified as quasars in SDSS DR9, with the rest being galaxies, which generally have lower redshifts in the sample. Remarkably, there are three known cases of gravitational lenses in our list (J0134$-$0931, J0414+0534, and J1601+4316). For these sources, the redshifts of both the lensing objects and the lensed background sources are given in Table~\ref{table:outliers}. Two other AGNs (J1006+3454 and J1301+4634) may show double-peaked narrow [O\,{\sc III}] emission lines among the total of ten sources that have optical spectra available in the SDSS DR9. There are also a couple of objects with large extended radio structures or other interesting properties, which may well be related to the significant optical--radio positional differences. In what follows, we briefly review possible explanations of the large positional offsets, and discuss the properties of the most interesting individual sources in our sample.

\begin{table*}[h]
\caption[]{\label{table:outliers}List of the 51 optical--radio positional outlier objects.}
\begin{tabular}{l r r r r r c c c c}
\hline
\noalign{\smallskip}
Short designation & \multicolumn{2}{c}{Position in radio}& \multicolumn{3}{c}{Offset in optical} & $m_{\rm r}$ & $z$ & Type & Ref.\\
\noalign{\smallskip}
 & \multicolumn{1}{c}{RA (J2000.0)} & \multicolumn{1}{c}{Dec (J2000.0)} & \multicolumn{1}{c}{Total} & \multicolumn{1}{c}{RA} & \multicolumn{1}{c}{Dec} & & (lens, source) & & $z$\\
\noalign{\smallskip}
 & \multicolumn{1}{c}{h \quad m \quad s} & \multicolumn{1}{c}{\degr \quad \arcmin \quad \arcsec} & \multicolumn{1}{c}{mas} & \multicolumn{1}{c}{mas} & \multicolumn{1}{c}{mas} & mag & & & \\
\noalign{\smallskip}
\hline
\noalign{\smallskip}
J0038$-$2120       & 00 38 29.9547 & $-$21 20 04.023 & 225 & 117 & 192 & 18.59 & 0.338 & G & 1\\
J0041$+$1339       & 00 41 17.2110 & 13 39 27.527 & 348 & 179 & $-$298 & 22.53 &  & G & \\
J0106$+$3402       & 01 06 00.2934 & 34 02 02.988 & 431 & 428 & $-$52 & 24.05 & 0.579 & G & 2\\
J0106$+$2539       & 01 06 10.9690 & 25 39 30.496 & 353 & $-$119 & $-$333 & 17.41 & 0.199 & G & 2\\
J0106$-$0315       & 01 06 43.2287 & $-$03 15 36.296 & 207 & 44 & $-$202 & 17.65 &  & Q & \\
\noalign{\smallskip}
J0134$-$0931$\ast$ & 01 34 35.6666 & $-$09 31 02.879 & 175 & 36 & $-$172 & 21.35 & (0.765, 2.216) & G,Q & 3, 4\\
J0146$+$2110       & 01 46 58.7839 & 21 10 24.381 & 277 & 122 & $-$248 & 23.03 &  & Q & \\
J0216$-$0118       & 02 16 05.6638 & $-$01 18 03.397 & 193 & $-$162 & $-$106 & 18.73 &  & Q & \\
J0216$-$0105       & 02 16 12.2119 & $-$01 05 18.826 & 179 & $-$141 & $-$111 & 17.68 & 1.492 & G & 5\\
J0334$+$0800       & 03 34 53.3167 & 08 00 14.419 & 272 & $-$179 & $-$205 & 22.49 & 1.982 & G & 2\\
\noalign{\smallskip}
J0335$-$0709       & 03 35 57.0552 & $-$07 09 55.854 & 347 & 329 & 109 & 21.86 &  & G & \\
J0414$+$0534$\ast$ & 04 14 37.7678 & 05 34 42.335 & 189 & $-$189 & $-$1 & 22.72 & (0.958, 2.639) & G,Q & 6, 7\\
J0431$+$2037       & 04 31 03.7614 & 20 37 34.265 & 203 & 168 & 114 & 18.04 & 0.219 & G & 8\\
J0435$+$2532       & 04 35 34.5829 & 25 32 59.697 & 249 & $-$238 & 75 & 20.25 &  & Q & \\
J0523$+$6007       & 05 23 11.0082 & 60 07 45.720 & 194 & $-$43 & 189 & 23.02 &  & Q & \\
\noalign{\smallskip}
J0552$+$0313       & 05 52 50.1015 & 03 13 27.243 & 281 & $-$220 & 175 & 23.14 &  & G & \\
J0644$+$2911       & 06 44 44.8158 & 29 11 04.018 & 264 & $-$176 & 197 & 21.84 &  & G & \\
J0729$-$1320       & 07 29 17.8177 & $-$13 20 02.272 & 448 & 360 & $-$266 & 16.68 &  & G & \\
J0736$+$2954       & 07 36 13.6611 & 29 54 22.186 & 192 & 168 & 94 & 22.49 &  & Q & \\
J0817$+$3227       & 08 17 28.5423 & 32 27 02.926 & 226 & $-$85 & $-$210 & 21.08 &  & G & \\
\noalign{\smallskip}
J0843$+$4537       & 08 43 07.0942 & 45 37 42.897 & 177 & 163 & $-$71 & 17.57 & 0.192 & G & 5\\
J0854$+$6218       & 08 54 50.5763 & 62 18 50.191 & 176 & $-$115 & $-$134 & 18.26 & 0.267 & G & 5\\
J0902$+$4310       & 09 02 30.9200 & 43 10 14.166 & 228 & $-$201 & 108 & 20.57 & 2.41 & Q & 9\\
J1006$+$3454$\ast$ & 10 06 01.7503 & 34 54 10.401 & 217 & $-$180 & 120 & 15.09 & 0.099 & G & 5\\
J1022$+$4239       & 10 22 13.1323 & 42 39 25.612 & 260 & 259 & $-$23 & 22.59 & 0.991 & G & 2\\
\noalign{\smallskip}
J1033$+$3935       & 10 33 22.0610 & 39 35 51.083 & 259 & $-$259 & 14 & 21.45 & 1.095 & G & 10\\
J1150$+$4332       & 11 50 16.6027 & 43 32 05.906 & 202 & 183 & $-$86 & 20.94 & 3.037 & Q & 11\\
J1254$+$0859       & 12 54 58.9577 & 08 59 47.549 & 239 & $-$230 & $-$65 & 23.00 &  & G & \\
J1301$+$4634$\ast$ & 13 01 32.6063 & 46 34 02.940 & 202 & 130 & $-$154 & 16.33 & 0.206 & G & 5\\
J1312$+$2531       & 13 12 14.2889 & 25 31 13.175 & 336 & 180 & 284 & 23.32 &  & G & \\
\noalign{\smallskip}
J1312$+$4828       & 13 12 43.3537 & 48 28 30.941 & 411 & 395 & $-$112 & 20.95 & 0.501 & G & 2\\ 
J1313$+$6735       & 13 13 27.9863 & 67 35 50.382 & 439 & $-$1 & $-$439 & 21.92 &  & G & \\
J1414$+$4554       & 14 14 14.8526 & 45 54 48.720 & 171 & $-$164 & 47 & 20.16 & 0.186 & G & 11\\
J1440$+$0127       & 14 40 33.6470 & 01 27 05.210 & 262 & 262 & 1 & 21.06 &  & G & \\
J1451$+$1343       & 14 51 31.4910 & 13 43 24.001 & 420 & 234 & 349 & 21.86 &  & G & \\
\noalign{\smallskip}
J1456$+$5048       & 14 56 08.1197 & 50 48 36.300 & 228 & 212 & $-$85 & 22.87 & 0.480 & Q & 12\\
J1503$+$0917       & 15 03 00.8995 & 09 17 58.983 & 291 & $-$290 & 24 & 21.98 &  & G & \\
J1506$+$3730$\ast$ & 15 06 09.5300 & 37 30 51.133 & 316 & $-$237 & $-$209 & 21.23 & 0.672 & G & 13\\
J1526$-$1351$\ast$ & 15 26 59.4407 & $-$13 51 00.164 & 192 & $-$78 & $-$176 & 19.44 & 1.687 & Q & 14\\
J1543$+$0452       & 15 43 33.9258 & 04 52 19.320 & 413 & 409 & $-$58 & 13.60 & 0.040 & G & 5\\
\noalign{\smallskip}
J1601$+$4316$\ast$ & 16 01 40.5154 & 43 16 46.477 & 322 & $-$139 & 291 & 20.84 & (0.414, 1.589) & G,Q & 15\\
J1603$+$1554       & 16 03 38.0619 & 15 54 02.355 & 205 & 87 & 185 & 15.22 & 0.110 & G & 5\\
J1604$+$1926       & 16 04 49.9938 & 19 26 20.942 & 194 & 109 & $-$160 & 22.36 &  & G & \\
J1625$+$4134       & 16 25 57.6697 & 41 34 40.629 & 259 & 109 & 235 & 22.35 & 2.55 & G & 14\\
J1648$+$2224       & 16 48 01.5356 & 22 24 33.148 & 179 & $-$23 & 178 & 21.70 & 0.823 & G & 5\\
\noalign{\smallskip}
J2052$+$1619       & 20 52 43.6199 & 16 19 48.828 & 231 & 229 & 31 & 21.82 &  & Q & \\
J2150$+$1449       & 21 50 23.6071 & 14 49 47.895 & 189 & $-$38 & 185 & 21.84 &  & Q & \\
J2210$+$0857       & 22 10 06.0503 & 08 57 29.564 & 193 & $-$86 & 173 & 18.84 &  & G & \\
J2259$-$0811       & 22 59 00.6888 & $-$08 11 03.043 & 235 & $-$231 & $-$41 & 20.32 & 1.380 & Q & 5\\
J2346$+$3011       & 23 46 46.2508 & 30 11 59.249 & 285 & 122 & 258 & 22.88 &  & Q & \\
\noalign{\smallskip}
J2347$-$1856       & 23 47 08.6267 & $-$18 56 18.858 & 244 & $-$205 & $-$134 & 22.68 &  & G & \\
\noalign{\smallskip}
\hline
\end{tabular}
\tablebib{(1)~\citet{mccarthy}; (2) \citet{healey}; (3) \citet{hall}; (4) \citet{gregg}; (5) \citet{ahn}; (6) \citet{tonry}; (7) \citet{lawrence}; (8) \citet{wright}; (9) \citet{hook}; (10) \citet{vermeulen}; (11) \citet{falco}; (12) \citet{bade}; (13) \citet{carilli}; (14) \citet{hewitt}; (15) \citet{fassnacht}.}
\end{table*}


\section{Discussion}
\label{sec:discussion}

When comparing directly the radio and optical positions of AGNs, we naturally assume that the optical and radio emission peaks physically and spatially coincide. Because the activity of these distant extragalactic sources is driven by matter accretion onto their central supermassive black holes and is confined to their close vicinity, this seems a plausible first approximation. Theoretically, the apparent origin of the inner radio jet, usually called the VLBI core, depends on the observing frequency due to opacity effects \citep[see e.g.][]{lobanov}. Actual VLBI measurements of this core shift indicate that it can be as high as $\sim$1~mas for certain sources \citep[e.g.][]{kovalev,sokolovsky}. \citet{kovalev} estimate the average shift between the cm-wavelength radio core and the optical core as $\sim$0.1~mas. This is more than two orders of magnitude lower than the SDSS positional accuracy. The core shift is therefore negligible and cannot be the cause of large optical--radio outliers in our sample. However, this effect should be considered for the alignment of the future {\em Gaia} optical reference frame with the ICRF. Similar arguments are valid for possible optical photocentric variability of quasars \citep{popovic}: while the effect can be as large as several mas for low-redshift AGNs, it is certainly negligible in our case.
In the next subsections, we look into some causes that could possibly be behind the positional outliers we found.

\subsection{Errors in positions}

Before discussing possible astrophysical explanations, we investigate whether systematic astrometric calibration errors in the SDSS database can cause the occasionally large optical--radio positional offsets. In the course of our work, we performed similar analyses to find optical counterparts to ICRF2 radio sources by cross-referencing their positions with the SDSS DR7 \citep{abazajian} and DR8 \citep{aihara1}, and calculated the optical--radio coordinate differences. The SDSS DR7 catalogue covers $\sim$12\,000 square degrees in the sky, mainly in the northern Galactic hemisphere and around the equator. Its astrometric calibration is similar and the accuracy is basically identical with DR9 \citep{pier}, apart from some proper motion errors for stars at low Galactic latitudes later corrected in DR9. Among the 51 positional outlier sources we found in DR9 (Table~\ref{table:outliers}), 37 are located and identified as photometric objects in DR7 as well, reflecting its smaller sky coverage. We used this overlapping subsample to examine the consistency of outliers between the two databases, and found that 27 of the 37 sources are outliers in both DR7 and DR9. None of them is associated with the reportedly miscalibrated runs in DR7 \citep{ahn}. This 73\% overlap between the two lists suggests that although the majority of our outliers listed in Table~\ref{table:outliers} are robust detections, there might be cases where small (local) astrometric calibration issues in SDSS contribute to the positional offsets. Worth noting is that the optical counterparts of ICRF2 radio sources are typically faint ($r$=18.88$\pm$1.83~mag), close to the SDSS limiting magnitude, therefore their optical positions are less precisely determined. These problems will be alleviated when the more accurate {\em Gaia} catalogue becomes available for such an optical--radio study.

We repeated our analysis with DR8, which has the same sky coverage as DR9. However, when calculating the offsets between the optical and radio coordinates, we found an anomaly in the declination differences. Almost all sources above $\delta$$\approx$$40\degr$ consistently showed an offset of $\sim$260~mas relative to the ICRF2 coordinates. This indicated a systematic error in the astrometric calibration of SDSS DR8. Our independent result is consistent with what is reported and detailed by \citet{aihara2}. This astrometric calibration problem has been fixed in the SDSS, and also prompted more rigorous astrometric quality-assurance measures in DR9 using a set of reference catalogues \citep{ahn}. This justifies the use of DR9 for our study. According to our normal probability plots in Fig.~\ref{img:qqplot}, there are no dominant systematic components in the optical--radio offsets using DR9. (The same is valid for DR7.) This means that there are no global errors in the astrometric reductions of these data releases. The even distribution of the outlier AGNs in the DR9 sky coverage (Fig.~\ref{img:skyplot}) also supports this notion.

\subsection{Errors in identifications}

As we have shown in Sect.~\ref{subsec:sourceselection}, our source selection method with a 500~mas search radius around the positions of ICRF2 objects guarantees that the matched optical--radio AGN sample is practically free from false identifications, i.e. physically unrelated radio and optical sources. This is because the average distance between quasars in the SDSS is much larger, in the order of arcminutes \citep[cf.][]{palanque}. However, it is in principle not excluded that the radio position of a particular object does not refer to the AGN core (which is in fact the base of the inner jet, close to the central engine) but to a brighter and compact component in one of the outward moving relativistic radio jets.
This is certainly not the case for the ICRF2 sources in general, since these are among the most prominent compact radio AGNs, often with sensitive multi-frequency radio imaging observations. But for at least one of our outlier sources, a peculiar quasar with a complex extended two-sided jet structure (J1526$-$1351), the ICRF2 catalogue indeed contains the position of a jet component  is actually brighter than the quasar core (see Sect.~\ref{subsec:individualsources} for the details).

\subsection{Gravitational lensing}

There could be astrophysical explanations for optical--radio offsets of $\sim$100~mas as well. Strong lensing caused by the intervening gravitational potential of a foreground galaxy can produce multiple and distorted images of a background object. The separations of gravitationally lensed AGN images are typically in the order of $1\arcsec$ or less. Thus the optical images are blended, possibly together with the lensing galaxy, and this complex structure remains unresolved in the SDSS. The optical position refers to the photocentre. On the other hand, the radio coordinates derived from VLBI data with much higher angular resolution usually refer to a particular (the brightest and most compact) image of the gravitationally lensed source. A special scenario where the lensed quasar is radio-loud but optically faint while the lensing galaxy is optically bright cannot be excluded. This could also lead to sub-arcsec apparent separation between the optical and radio positions.

Successful systematic searches for gravitational lenses in the SDSS database were performed using spectroscopic data \citep[e.g.][]{bolton} or morphological and colour-selection criteria \citep[e.g.][]{inada}. The candidates were then followed-up by higher-resolution imaging and spectroscopic observations for verification. Recent simulations show that {\em Gaia} will detect $\sim$0.6\% of the quasars that are multiply imaged due to gravitational lensing \citep{finet}. Here we deal with radio-selected AGNs taken from the ICRF2 catalogue. To estimate the probability of finding lensed objects in our case, we therefore applied the result from the largest sample of strongly lensed flat-spectrum radio sources in the Cosmic Lens All-Sky Survey \citep[CLASS,][]{browne}, where $\sim$0.14\% was found for the point-source lensing rate. In sharp contrast with the statistical expectation, 3 of our 51 outlier sources ($\sim$6\%) are known gravitational lens systems (see Sect.~\ref{subsec:individualsources} for details). This indicates that the method of finding AGNs with significant optical--radio positional offsets may be an efficient tool for identifying gravitational lens candidates. Note that for many of the objects in Table~\ref{table:outliers}, optical spectroscopic data are unavailable at present, and sub-arcsecond separation gravitational lensing cannot be convincingly excluded.

\subsection{Interacting active galactic nuclei}

Another possible astrophysical cause for the measured optical--radio positional offsets could be related to dual AGNs. The presence of dual accreting supermassive black hole systems is a natural consequence of hierarchical structure formation in the Universe through mergers of galaxies \citep[e.g.][]{begelman}. Hydrodynamical simulations \citep{vanwassenhove} suggest that simultaneous AGN activity is mostly expected at the late stages of mergers, at or below $\sim$1--10~kpc separations. Assuming a cosmological model with $H_{\rm 0}$=70~km~s$^{-1}$~Mpc$^{-1}$, $\Omega_{\rm m}$=0.3, and $\Omega_{\Lambda}$=0.7, 1~kpc linear size corresponds to $\la$300~mas angular size at any plausible redshift above $z$=0.2, or smaller if the dual system is inclined to the plane of the sky. Thus the expected angular separations are typically below the resolution limit of SDSS, but comparable to the optical--radio positional offsets found in this paper. For example, a dual AGN system of which one component is presently in its short-lived radio-loud activity phase but the other one is not, would easily result in a detectable positional offset.

The SDSS spectroscopic database was used to find candidate dual AGNs via searching for double-peaked narrow [O {\sc III}] or other emission line profiles \citep[e.g.][]{smith}. These may indicate gravitationally bound dual AGNs with distinct narrow-line regions (NRL). In this model, the two components have different radial velocities due to their orbital motion, resulting in the doubling of the line profiles. However, peculiar gas kinematics and jet--cloud interaction in a single NLR can also lead to similar spectral signatures. Indeed, confirmed kpc-scale dual AGNs seem to add up only a few percent of the candidates with double-peaked [O {\sc III}] emission lines \citep{shen}. According to the estimates of \citet{rosario} and \citet{fu}, only $\sim$0.3\% or less of the low-redshift ($z$$\la$0.6) SDSS quasars host dual accreting black holes separated on kpc scales. This ratio decreases with increasing redshifts \citep{yu}.

A sample of objects with large optical--radio positional offsets {\em and} double-peaked [O {\sc III}] optical emission lines may provide better candidates for actual dual AGNs. In our case, only 10 out of 51 sources have optical spectra available in the SDSS DR9. By visual inspection, two of them show indication of double peaks in their [O {\sc III}] emission lines (see Sect.~\ref{subsec:individualsources}), albeit with a low signal-to-noise ratio. With more complete spectral data, more precise {\em Gaia} astrometry in the future, and additional observational verifications \citep[e.g. with adaptive optics imaging or optical slit spectroscopy,][]{shen,rosario}, this possibility is worth investigating in more detail.

After final coalescence, a supermassive black hole may be kicked out of the centre of its host galaxy. Numerical simulations of recoiling black holes predict up to kpc-scale separations \citep[for a recent review, see][]{komossa}. Observationally, these objects could also appear as radio AGNs offset from their host galaxies.

\subsection{Comments on individual sources}
\label{subsec:individualsources}

Properties that may be relevant for this study are collected from the literature for the following individual objects in Table~\ref{table:outliers}.  

{\bf J0134$-$0931.}
This object is the quintuple quasar, a gravitationally lensed compact radio source at $z_{\rm s}$=2.216 consisting of five components with a maximum separation of $\sim$$0\farcs7$ \citep{winn1,gregg,hall,winn2,keeton}. The VLBI coordinates refer to one of the lensed images, component A. On the other hand, all five images of the background quasar, as well as the lensing pair of galaxies ($z_{\rm l}$=0.765), are unresolved in SDSS. It is therefore not surprising that we found an offset of 175~mas between the radio position of component A and the SDSS optical position, which should refer to the peak of the blended emission of the lensing galaxy pair and the gravitationally lensed components. 

{\bf J0414+0534.}
This radio source at $z_{\rm s}$=2.639 \citep{lawrence}, also known as 4C~+05.19, is gravitationally lensed \citep{hewitt2} by an elliptical galaxy \citep{schechter} at $z_{\rm l}$=0.958 \citep{tonry}. The system of four strong lensed radio components separated by nearly $2\arcsec$ has been the subject of several VLBI studies \citep[e.g.][]{patnaik,trotter,ros,volino}. A partial Einstein ring connecting the three brightest images is seen in a deep high-resolution {\em Hubble Space Telescope} (HST) image \citep{falco2}. The ICRF2 coordinates refer to component A1, while the optical position is affected by the complex structure of the different quasar images and the lens, only partially resolved in SDSS.   

{\bf J1006+3454.}
Also known as 3C~236, this prominent radio source at $z$=0.1 is among the largest radio galaxies observed \citep[e.g.][and references therein]{willis,barthel,schilizzi,labiano}. The overall radio structure of this extensively studied object has a narrow morphology of $\sim$$40\arcmin$ angular size, extending in about SW--NE position angle. Radio interferometric observations reveal a double--double morphology, with a 2-kpc scale compact steep-spectrum (CSS) source inside. The accurate VLBI position refers to the component B2 identified with the radio core by \citet{schilizzi}. The SDSS optical position is offset by 217~mas, roughly along the position angle of the radio structure. The optical--radio offset is consistent with Fig.~8 of \citet{odea} that displays an overlay of VLBI and HST images. The apparently renewed radio activity in 3C~236, the presence of young star-forming regions in the dust lane \citep{odea}, and the disturbed kpc-scale optical morphology suggest a galaxy merger that might have led to the reignition of the radio AGN \citep[see e.g.][and references therein]{labiano}. Notably, the SDSS DR9 spectrum of J1006+3454 (3C~236) hints at a double-peaked narrow 5007~$\AA$ [O\,{\sc III}] emission line.

{\bf J1301+4634.}
A low-redshift ($z$=0.2) galaxy, which is the central, brightest member of a small cluster selected from SDSS photometric data \citep{koester}. Visual inspection of the SDSS DR9 spectrum suggests broadened or double-peaked narrow [O\,{\sc III}] emission lines. 

{\bf J1506+3730.}
Although the radio-loud AGNs are typically found in ellipticals, the host of J1506+3730 is an inclined disk galaxy, with an optically obscured nucleus. Significant neutral hydrogen and molecular absorption is detected towards the radio AGN, arising from a fast gas outflow, a possible result of jet--cloud interaction \citep[e.g.][and references therein]{carilli2,kanekar}. The optical--radio positional difference we found can be reconciled with the fact that the nucleus of this red quasar is heavily obscured in the optical. The SDSS DR9 optical position is offset from the VLBI position by 316~mas in the $\sim$SW direction, broadly coinciding with the position angle of the radio jet \citep[e.g.][]{polatidis}. 

{\bf J1526$-$1351.}
This is a high-luminosity CSS quasar, with a complex radio structure spanning $\sim$$1\arcsec$. Based on polarization-sensitive and dual-frequency VLBI imaging observations, \citet{mantovani1,mantovani2} found a peculiar non-collinear two-sided jet in this source. The flat-spectrum core is $\sim$100 mas south of the radio peak, whose position is listed in the ICRF2 catalogue. The peak in fact coincides with a bright and compact component at the end of the northern jet. It is therefore reasonable to assume that the optical AGN position is closer to the radio core, roughly explaining the offset we found. 

{\bf J1601+4316.}
This source is the third case of known strong gravitational lenses in our sample, discovered by \citet{jackson}. The lens is a spiral galaxy at $z_{\rm l}$=0.414, the background source is a doubly imaged quasar at $z_{\rm s}$=1.589 \citep{fassnacht}. The ICRF2 radio catalogue lists the coordinates of both the brighter component A and the weaker component B, separated by $1\farcs4$ \citep[e.g.][]{koopmans,patnaik2}. Our selection algorithm found an optical counterpart to component B in SDSS DR9, the one seen very close to the lens \citep{jaunsen}. The position angle of our optical--radio positional offset (322~mas towards $\sim$NW) is consistent with the direction of the lensing galaxy and component A. 


\section{Conclusions and outlook}
\label{sec:conclusion}

We performed a case study for directly linking the radio and optical reference frames using common objects in the ICRF2 \citep{fey2} and SDSS DR9 \citep{ahn} catalogues. We found optical counterparts for 1297 radio-loud AGNs, practically all that are located in the SDSS DR9 footprint in the sky, within a search radius of $0\farcs5$ around the accurate VLBI-determined positions. Our overlapping ICRF2--SDSS DR9 sample is free from false identifications. The optical--radio coordinate differences in right ascension and declination follow normal distributions with fat tails. This allows us to characterize the overall astrometric accuracy of SDSS DR9: both equatorial coordinates of the matching extragalactic objects are in general consistent with no offset with respect to the ICRF2, with $\sim$55~mas standard deviation. 

We identified 51 AGNs ($\sim$4\% of the sample) for which the optical--radio positional offset is significant, exceeding 170~mas ($\sim$3$\sigma$). We argued that there is an astrophysical cause behind the majority of these outliers. The presence of significant positional outliers underlines the importance of a careful reference source selection for the precise alignment of the most accurate radio reference frame with the future optical frame to be constructed from the measurements of the {\em Gaia} astrometric space telescope. 

The question of optical--radio positional differences was also investigated by \citet{camargo} using a different approach, with a smaller sample. They performed accurate relative optical astrometric measurements in the fields around 22 ICRF2 sources. Four of their sources showed offsets relative to their ICRF2 positions larger than their 3$\sigma$ confidence level, $\sim$80~mas. \citet{camargo} concluded that these separations cannot be explained merely by statistical fluctuations or systematic errors in the optical reference frame, and they might be related to the relatively more complex VLBI structure of the given quasars.  

We offered some viable explanations for the positional outliers listed in Table~\ref{table:outliers}. As shown for the peculiar quasar J1526$-$1351 as an example, the ICRF2 position may refer to a bright component farther along the radio jet, and not to the true VLBI core, which is supposed to be close to the location of the central supermassive black hole. Although this situation is believed to be rare if not unique, firm proof should come from sensitive, high-resolution multi-frequency VLBI observations of individual radio reference frame objects, not just typical snapshot imaging. 

There are three known cases of strong gravitational lensing in our sample, nearly 50 times more than expected on statistical grounds. For these objects, the $\sim$100-mas scale offsets between the radio position (which refers to one of the lensed images) and the optical photocentre is not surprising. It is possible that there are more gravitationally lensed objects in our sample, waiting for identification by follow-up spectrosopic observations or sensitive high-resolution imaging.  

Positional non-coincidences between the optical and the radio can also be caused by interacting AGNs containig dual AGNs. Observationally, known dual AGNs with $\sim$1--10-kpc scale separations are rare, partly because the period of their simultaneous activity within the lifetime of the AGNs is probably short. Moreover, it is difficult to confirm their existence due to the high-resolution and sensitive imaging and spectroscopic observations required. At present, there is no efficient selection method to apply for finding firm dual AGN candidates. It remains to be seen if the significant optical--radio positional difference in the case of some radio-loud AGNs found in this paper is a good indication of AGN duality. A much larger sample will be offered for such a study by the {\em Gaia} astrometric catalogue of extragalactic sources. Once this becomes available, as \citet{browne2} suggests, a future extensive radio interferometric survey conducted with the e-MERLIN array in the United Kingdom, targeting elliptical galaxies that host radio-loud AGNs, could reveal displacements. These may arise either from dual AGNs or recoiling supermassive black holes that are expelled from the galaxy after the coalescence of a binary system \citep{komossa}. On a short term, sensitive e-MERLIN imaging of our sample in Table~\ref{table:outliers}, probing the $\sim$100-mas scale radio stucture, should be able to provide clues for tracking down the causes of significant positional offsets between SDSS DR9 and ICRF2. Looking for signatures of gravitational lensing or double-peaked narrow optical emission lines would also require high-quality spectroscopic observations of these optically faint sources.       


\begin{acknowledgements}

This work is supported by the Hungarian Scientific Research Fund (OTKA K104539) and the grant T\'AMOP-4.2.2.B-10/1--2010-0009.
This research has made use of the NASA/IPAC Extragalactic Database (NED) which is operated by the Jet Propulsion Laboratory, California Institute of Technology, under contract with the National Aeronautics and Space Administration.
Funding for the SDSS-II and SDSS-III has been provided by the Alfred P. Sloan Foundation, the Participating Institutions, the National Science Foundation, the U.S. Department of Energy Office of Science, the National Aeronautics and Space Administration, the Japanese Monbukagakusho, the Max Planck Society, and the Higher Education Funding Council for England. The SDSS web site is \url{http://www.sdss.org/}, the SDSS-III web site is \url{http://www.sdss3.org/}, where the list of the other participating and collaborating institutions can be found.

      
\end{acknowledgements}



\end{document}